\documentstyle[eqsecnum,preprint,aps,prd,epsfig]{revtex}
\date{\today}
\tighten
\begin{document}
\draft
\def\diag{\mathop {\rm diag}}

%%%%%%%%%%%%%%%%%%%%%%%%%%%%%%%%%%%%%%%%%%%%%%%%%%%%%%%%%%%%%%%%%%%%%%%%%%%%%%%

\title{WAVEFORMS FOR GRAVITATIONAL RADIATION \\ FROM COSMIC STRING LOOPS}

\author{Bruce Allen}
\address{Department of Physics, University of Wisconsin -- Milwaukee\\
P.O. Box 413, Milwaukee, Wisconsin 53201, U.S.A.\\
email: ballen@dirac.phys.uwm.edu}

\author{Adrian C. Ottewill}
\address{Department of Mathematical Physics\\
University College Dublin, Belfield, Dublin 4, IRELAND\\
email: ottewill@relativity.ucd.ie}

\maketitle
\begin{abstract}

We obtain general formulae for the plus- and cross- polarized waveforms of
gravitational radiation emitted by a cosmic string loop in transverse,
traceless (synchronous, harmonic) gauge. These equations are then
specialized to the case of piecewise linear loops, and it is shown
that the general waveform for such a loop is a piecewise linear
function.
We give several simple examples of the waveforms from such loops. 
We also discuss the relation between the gravitational radiation 
by a smooth loop and by a   piecewise linear approximation to it.

\end{abstract}
\pacs{PACS number(s): 98.80.Cq, 04.30.Db, 11.27.+d}

%%%%%%%%%%%%%%%%%%%%%%%%%%%%%%%%%%%%%%%%%%%%%%%%%%%%%%%%%%%%%%%%%%%%%%%%%%%%%%%

\section{INTRODUCTION}
\label{introduction}

Cosmic strings are one dimensional topological defects that may have
formed if the vacuum underwent a phase transition at very early times
breaking a local $U(1)$ symmetry
\cite{Kibble,Zel'dovich,Vilenkin,SV}.  The resulting network of strings
is of cosmological interest if the strings have a large enough mass per
unit length, $\mu$.  If $G\mu/c^2\sim 10^{-6}$, where $G$ is
Newton's constant and $c$ is the speed of light (i.e. $\mu \sim
10^{22}$g/cm) then cosmic strings may be massive enough to have
provided the density perturbations necessary to produce the large scale
structure we observe in the Universe today and could
explain the pattern of anisotropies
observed in the cosmic microwave background \cite{AllenPRL}. 

The main constraints on $\mu$ come from observational bounds on the
amount of gravitational background radiation emitted by cosmic string
loops ($\!$\cite{SV,AllenCaldwell,Caldwell} and references therein).  A
loop of cosmic string is formed when two sections of a long string (a
string with length greater than the horizon length) meet and
intercommute.  Once formed, loops begin to oscillate under their own
tension, undergoing a process of self-intersection (fragmentation) and
eventually creating a family of non-self-intersecting oscillating
loops.  The gravitational radiation emitted by each loop as it
oscillates contributes to the total background gravitational
radiation.

In a  pair of papers, we introduced and tested a new method for
calculating the rates at which energy and momentum are radiated by
cosmic strings \cite{AC,ACO2}. 
 Our investigation found that many of the published
radiation rates were numerically inaccurate (typically too low by a
factor of two). Remarkably, we also found a lower bound 
(in the center-of-mass frame)  for the rate of gravitational radiation
from a cosmic string loop \cite{AC2}.
Our method involved the use of piecewise linear cosmic strings.  In
this paper we wish to provide greater insight into the behaviour of
such loops and, in particular, how they approximate smooth loops
by examining the waveforms of the gravitational waveforms of such loops.

It has long been known \cite{AllenCaldwell,LesHouches} that the first generation of ground-based interferometric
gravitational-wave detectors (for example, LIGO-I) will not be able to
detect the gravitational-wave stochastic background produced by a network
of cosmic strings in the universe. The amplitude of this background is
too weak to be detectable, except by a future generation of more advanced
instruments. However, a recent paper by Damour and Vilenkin \cite{DamourVilenkin} has shown
that the non-Gaussian bursts of radiation produced by cusps on the closest
loops of strings would be a detectable LIGO-I source. 
While the specific examples studied here do not include these types
of cusps the general method developed can be applied to such loops.

Our space-time conventions follow those of Misner, Thorne and Wheeler 
\cite{MTW} so that $\eta_{\mu\nu} = \diag(-1,1,1,1)_{\mu\nu}$.  We
also set $\hbar = c =1$, but we leave $G$ explicit.

\section{GENERAL THEORY}
\label{theory}
In the center-of-mass frame, a
cosmic string loop is specified by the 3-vector position ${\bf x}(t,\sigma)$
of the string as a function of two variables: time $t$ and a space-like
parameter $\sigma$ that runs from $0$ to $L$.  (The total energy of the
loop is $\mu L$.)
When the gravitational back-reaction is neglected, 
(a good approximation if $G\mu^2 <\!\!< 1$), the string loop
satisfies equations of motion whose most general solution in the
center-of-mass frame is
\begin{equation}
{\bf x}(t,\sigma)= {1 \over 2} \bigl[{ \bf a}(u) + {\bf b}(v)\bigr].
\label{X}
\end{equation}
where 
\begin{equation}
u = t+\sigma, \qquad v = t- \sigma.
\end{equation}
Here ${\bf a}(u) \equiv {\bf a}(u+L)$ and ${\bf b}(v) \equiv {\bf b}(v+L)$ are a pair of
periodic functions, satisfying the ``gauge condition"
$|{\bf a}'(u)| = |{\bf b}'(v)|=1$, where $'$ denotes differentiation
with
respect to the
function's argument. 
Because the functions ${\bf a}$ and ${\bf b}$ are periodic in their
arguments, the string loop is periodic in time.  The period of the
loop is $L/2$ since
\begin{equation}
{\bf  x}(t+{L\over 2},\sigma+{L\over 2})={1\over 2}[
{\bf a}(t+\sigma+L)+ {\bf  b}(t-\sigma)]={1\over 2}[{\bf a}(t+\sigma)+
{\bf b}(t-\sigma)]= {\bf x}(t,\sigma).
\label{periodicity}
\end{equation}

With our choice of coordinates and gauge, the energy-momentum tensor
$T^{\mu\nu}$ for the string loop is given  by 
\begin{equation}
T^{\mu\nu}(t, {\bf y})=\mu \mathop{\int\!\int}\limits_{0 \leq u-v \leq 2L}
du\, dv\>
G^{\mu\nu}(u,v)  \delta\bigl( t - {\textstyle{1 \over 2}} (u+v) \bigr)
\delta^{(3)}\bigl({\bf y} - {\bf x}(u,v)\bigr),\label{tmunu}
\end{equation}
where $G^{\mu\nu}$ is defined by
\begin{equation}
G^{\mu\nu}(u,v)=\partial_ux^\mu\partial_vx^\nu+
\partial_vx^\mu\partial_ux^\nu,
\label{Gmunu}
\end{equation}
with $x^\mu=\bigl(t,{\bf x}(t,\sigma)\bigr)$.
In terms of ${\bf a}$ and ${\bf b}$,
\begin{equation}
  G^{00} = {\textstyle{1 \over 2}} , 
\qquad
 G^{0i} = {\textstyle{1 \over 4 }} [a'_i + b'_i]  ,
\qquad
 G^{ij} = {\textstyle{1 \over 4 }} [a'_i b'_j + a'_j b'_i] ,
\end{equation}
and the trace is
\begin{equation}
 G^\lambda{}_{\lambda}  = {\textstyle{1 \over 2 }} [ -1 + {\bf
   a}'{\cdot}{\bf b}']  .
\end{equation}
Alternatively we may introduce the four-vectors 
$A^\mu(u) = \bigl(u, {\bf a}(u)\bigr)$ and 
$B^\mu(v) = \bigl(v, {\bf b}(v)\bigr)$ so that
\begin{equation}
\label{Gdef}
   G^{\mu\nu} = {\textstyle{1 \over 4 }} (A'{}^\mu B'{}^\nu 
                            + B'{}^\mu A'{}^\nu )  .
\end{equation}
The ``gauge conditions''  are satisfied if and only if
$A'{}^\mu(u)$ and $B'{}^\mu(v)$ are null vectors.

As a consequence of the time periodicity of the loop the stress tensor can
be expressed as a Fourier series
\begin{equation}
T_{\mu\nu}(t,{\bf y}) = \sum\limits_{n=-\infty}^\infty \> e^{i\omega_n t }\>
\tilde T_{\mu\nu}(\omega_n,{\bf y}) ,
\end{equation}
where $\omega_n = 4\pi n /L$ and 
\begin{eqnarray}
\tilde T_{\mu\nu}(\omega_n,{\bf y}) &=& {2 \over L} \int\limits_0^{L/2}
{\rm d}t \> e^{- i\omega_n t } T_{\mu\nu}(t,{\bf y})\nonumber \\
    &=& {2\mu \over L} \int\limits_0^{L}  du  \int\limits_0^{L} 
      dv \> e^ {- {1 \over 2} i \omega_n (u+v) } G^{\mu\nu} (u,v) 
              \delta^{(3)}\bigl({\bf y} - {\bf x}(u,v)\bigr)  .
\label{Tmunu_tilde}
\end{eqnarray}

The retarded solution for the linear metric perturbation due to 
this source in harmonic gauge is \cite{Weinberg}
\begin{equation}
\label{metric}
  h_{\mu\nu}(t, {\bf x}) = 4G  \sum\limits_{n=-\infty}^\infty  \int {d^3{\bf y}
  \over |{\bf x} - {\bf y}|} \>
  \Bigl[\tilde T_{\mu\nu}(\omega_n,{\bf y}) -
    {\textstyle {1 \over 2}} \eta_{\mu\nu} \tilde 
T^\lambda{}_{\lambda}(\omega_n,{\bf y})
           \Bigr] \, e^{i\omega_n (t - |{\bf x} - {\bf y}|)}  .
\end{equation}
Far from the string loop center-of-mass
 the dominant behavior  is that of an outgoing
spherical wave given by
 \begin{equation}
\label{far_field_metric}
  h_{\mu\nu}(t, {\bf x}) = 4G \sum\limits_{n=-\infty}^\infty  {e^{i\omega_n(t-r)}
      \over r} \int  d^3{\bf y} \>
  \Bigl[\tilde T_{\mu\nu}(\omega_n,{\bf y}) -
    {\textstyle {1 \over 2}} \eta_{\mu\nu} \tilde T^\lambda{}_{\lambda}(\omega_n,{\bf y})
           \Bigr] \, e^{i \omega_n \hat {\bf \Omega}{\cdot} {\bf y}}  ,
\end{equation}
where $r= |x|$ and $\hat {\bf \Omega} = {\bf x}/r$ is a unit vector pointing away
from the source.
Inserting Eq. (\ref{Tmunu_tilde}) into Eq. (\ref{far_field_metric}) we find the
field far from a cosmic string loop is 
 \begin{equation}
\label{far_field_string_metric}
  h_{\mu\nu}(t, {\bf x}) = {8G\mu \over L} \sum\limits_{n=-\infty}^\infty {e^{i\omega_n(t-r)}
      \over r} \int\limits_0^L du  \int\limits_0^L  dv \>
  \Bigl[\tilde G_{\mu\nu}(u,v) -
    {\textstyle {1 \over 2}} \eta_{\mu\nu} \tilde G^\lambda{}_{\lambda}(u,v)
           \Bigr] \, e^ {- i \omega_n\bigl[ {1 \over
      2}(u+v) 
                -  \hat {\bf \Omega}{\cdot} {\bf x}(u,v)\bigr]}  .
\end{equation}
The $n=0$ term in this sum corresponds
to the static field
 \begin{equation}
\label{static_field}
  h^{\rm static}_{\mu\nu}(t, {\bf x}) = {8G\mu \over r  L}
      \int\limits_0^L  {\rm d}u  \int\limits_0^L
    {\rm d}v \>
  \Bigl[ G_{\mu\nu}(u,v) -
    {\textstyle {1 \over 2}} \eta_{\mu\nu}  G^\lambda{}_{\lambda}(u,v)
           \Bigr]  ,
\end{equation}
 \begin{equation}
   = {2 G \mu L \over r} (\eta_{\mu\nu} + 2\hat t_\mu \hat t_\nu )
   = {2 G M \over r} \delta_{\mu\nu} ,
\end{equation}
as appropriate to a object with mass $M$ as may be seen by comparison with 
the Schwarzschild metric in isotropic coordinates (see, for example,
Eq.~(31.22) of Ref.~\cite{MTW}).  We denote the
radiative part of the field by
 \begin{equation}
    h^{\rm rad}_{\mu\nu} =  h_{\mu\nu } -  h^{\rm static}_{\mu\nu }.
\end{equation}

We may rewrite Eq. (\ref{far_field_string_metric}) as
 \begin{equation}
   h_{\mu\nu}(t, {\bf x}) =  \sum\limits_{n=-\infty}^\infty 
e^{- i\omega_n k_\mu x^\mu} {\rm e}^{(n)}{}_{\mu\nu}(\hat {\bf \Omega})
\end{equation}
where $k^\mu = (1,\hat {\bf \Omega})$ is a null vector in the direction
of propagation and 
\begin{equation}
{\rm e}^{(n)}{}_{\mu\nu} = {8GM \over r L^2 } \int\limits_0^L du  \int\limits_0^L  dv \>
  \Bigl[ G_{\mu\nu}(u,v) -
    {\textstyle {1 \over 2}} \eta_{\mu\nu} G^\lambda{}_{\lambda}(u,v)
           \Bigr] \, e^ { i{1 \over
      2} \omega_n\bigl[k_\mu A^\mu(u) + k_\mu B^\mu(v)\bigr]}
\end{equation}
are polarization tensors.
From Eq. (\ref{Gdef}), it is clear that the polarization tensors may 
be written in terms of the fundamental integrals
\begin{equation}
   I^{(n)}{}^\mu = 
{1 \over L}\int\limits_0^L du \> A'{}^\mu(u)\, e^{ i{1 \over
      2} \omega_n k_\mu A^\mu(u)} ,
\end{equation}
and 
\begin{equation}
   J^{(n)}{}^\mu = {1 \over L}\int\limits_0^L dv \>  B'{}^\mu(v)\,
 e^{ i{1 \over
     2} \omega_n k_\mu B^\mu(v)} .
\end{equation}
In terms of these integrals 
\begin{mathletters}
\begin{equation}
{\rm e}_{00} = {2G M \over r  } \left[I_0 J_0
+ {\bf I}{\cdot}{\bf J} \right]
\end{equation}
\begin{equation}
{\rm e}_{0i} = {2GM \over r  } \left[ I_0 J_i + J_0
I_i \right]       
\end{equation}
\begin{equation}
{\rm e}_{ij} = {2GM \over r  } \left\{ \left[ I_i J_j + J_i
I_j \right]  + \delta_{ij} \left[I_0 J_0
- {\bf I}{\cdot}{\bf J} \right] \right\} , 
\end{equation}
\end{mathletters}
where we have dropped the superscript $(n)$ for clarity.

The harmonic gauge condition requires that the polarization tensors satisfy
$k^\mu{\rm e}_{\mu\nu}= {1 \over 2} k_\nu {\rm e}_{\mu}{}^\mu$.
 This is easily verified by  noting that $I^0 = \hat {\bf
\Omega}{\cdot}{\bf I}$ and  $J^0 = \hat {\bf
\Omega}{\cdot}{\bf J}$.   These equations  follow from the identity
\begin{equation}
 \int\limits_0^L du \> k_\mu A'{}^\mu(u) \, e^ {- i{1 \over
      2} \omega_n k_\nu A^\nu(u)} =
{2i \over \omega} \int\limits_0^L du \> {d\ \over du} 
    \, e^{-i {1 \over 2} \omega_n k_\nu A^\nu(u)} = 0,
\end{equation}
which is a consequence of periodicity, and the corresponding equation
for $B^\mu$.  
The harmonic gauge condition does not determine the gauge completely
and we are left with the freedom to make transformations of the form
\begin{equation}
\label{gauge}
 {\rm e}'{}_{\mu\nu} =  {\rm e}_{\mu\nu} + k_\mu\varepsilon_\nu
 + k_\nu\varepsilon_\mu   .
\end{equation}
If we make the choice
\begin{mathletters}
\begin{equation}
\varepsilon_0 =  {G M \over r  } \left[ I_0 J_0 + {\bf I}{\cdot}{\bf J} \right]    
\end{equation}
and 
\begin{equation}
\varepsilon_i = {G M \over r  } \left\{ \left[ I_0 J_0 + {\bf I}{\cdot}{\bf J}
\right] \Omega_i +2 \left[ I_0 J_i + J_0 I_i  \right] \right\}   
\end{equation}
\end{mathletters}
then
\begin{equation}
 {\rm e}'{}_{0\mu} = 0  .
\end{equation}
The spatial components are given by 
\begin{eqnarray}
 {\rm e}'{}_{ij} =  {2 G M \over r  } \bigl\{ [I_i J_j + I_j
 J_i] + \delta_{ij} [I_0 J_0 &-& {\bf I}{\cdot}{\bf J}] +
   \Omega_i \Omega_j [I_0 J_0 + {\bf I}{\cdot}{\bf J}] \nonumber \\
    &+&
  I_0[J_i \Omega_j + \Omega_i J_j ] + J_0 [ I_i \Omega_j + \Omega_i
 I_j] \bigr\}  ,
\end{eqnarray}
these satisfy the gauge conditions
\begin{equation}
 {\rm e}'{}_{\mu}{}^\mu = {\rm e}'{}_i{}^i = 0  ,
\end{equation}
and 
\begin{equation}
 {\rm e}'{}_j{}^i \Omega_i = 0  .
\end{equation}

If we perform a spatial rotation to coordinates $(x',y',z')$ where
$\hat {\bf \Omega}$ points along the $z'$-axis then we can write
\begin{equation}
 {\rm e}'{}_{i'j'}  = \left(\matrix{e_+&e_\times&0\cr
                                      e_\times&-e_+&0\cr
                                      0&0&0} \right)  ,
\end{equation}
where
\begin{mathletters}
\begin{equation}
e_+ =   {2 GM \over r  } [ I_{1'}J_{1'} - I_{2'}J_{2'} ]
\end{equation}
and 
\begin{equation}
e_\times =   {2 GM \over r } [ I_{1'}J_{2'} + I_{2'}J_{1'} ] ,
\end{equation}
\end{mathletters}
define two modes of linear polarization.  

In terms of the original basis we can write
\begin{mathletters}
\begin{equation}
  e_+ = {2 G M \over r  } ( \cos 2 \psi A_+ + \sin 2 \psi A_\times)
\label{e_plus}
\end{equation}
and 
\begin{equation}
  e_\times = {2 G M \over r  } ( - \sin 2 \psi A_+ + \cos 2 \psi A_\times)
\label{e_cross}
\end{equation}
\label{es}
\end{mathletters}
with
\begin{mathletters}
\begin{eqnarray}
  A_+ = &&\Bigl\{
  (\cos^2\theta  \cos^2 \phi - \sin^2\phi) I_1J_1 
 + ( \cos^2\theta \sin^2 \phi - \cos^2 \phi) I_2 J_2 
           + \sin^2 \theta I_3  J_3\nonumber \\
&&+ ( 1 + \cos^2\theta) \cos \phi \sin
  \phi( I_2 J_1 + I_1 J_2)
     - \sin\theta \cos \theta \cos \phi( I_3 J_1 + I_1 J_3) \nonumber \\
  && -  \sin \theta \cos \theta \sin \phi(I_2 J_3 + I_3
  J_2)           \Bigr\}
\label{A_plus}  \\
  A_\times = && \Bigl\{
   - 2 \cos \theta \sin \phi \cos \phi I_1J_1 
  + 2  \cos \theta  \sin \phi \cos \phi I_2 J_2 
   +  \cos \theta (2 \cos^2\phi -1)( I_2 J_1 + I_1 J_2)\nonumber \\
&& 
     + \sin\theta\sin\phi ( I_3 J_1 +  I_1 J_3)
   -  \sin \theta  \cos \phi (I_2 J_3 + I_3 J_2)
           \Bigr\}
\label{A_times}
\end{eqnarray}
\label{As}
\end{mathletters}
where $\theta$, $\phi$ and $\psi$ are the Euler angles defining the
orientation of the frame $(x',y',z')$ relative to the original frame
(our conventions follow those of Ref.~\cite{Arfken}).
The corresponding linearly polarized waveforms are then defined by
 \begin{equation}
   h^{\rm rad}_{+/\times}(t, {\bf x}) = 
\sum\limits_{n=1}^\infty 
\Bigl\{  e^{i 4 \pi n (t-r)/L } {\rm e}^{(n)}_{+/\times} +
           e^{- i 4 \pi n (t-r)/L } {\rm e}^{(n)*}_{+/\times}\Bigr\}  .
\label{waveformsum}
\end{equation}
Recall that $h^{\rm rad}$ is obtained from the full metric perturbation
$h$ by dropping the $n=0$ term which corresponds to the static 
(non-radiative) part of the field.

The power emitted to infinity per solid angle may be written as 
 \begin{eqnarray}
     {d P \over d \Omega} &=& \lim_{r \to \infty} {r^2 \over 32 \pi G} \langle
     h_{\alpha\beta,t}  h^{\alpha\beta}{}_{,r} \rangle \\
   &=&  {GM^2 \over 2 \pi}\sum\limits_{n=1}^\infty   \omega_n^2
        \bigl\{ \langle|A_+^{(n)}|^2\rangle
            + \langle|A_\times^{(n)}|^2\rangle \bigr\}   .
\label{power}
\end{eqnarray}

\section{EXAMPLES}
\label{examples}

For convenience we shall now set the length of the loop $L =1$, and
take $\psi=0$.

\subsection{Piecewise linear loops}

These are loops for which the functions ${\bf a}(u)$ and ${\bf b}(v)$
are piecewise linear functions.  The functions ${\bf a}(u)$ and ${\bf
b}(v)$
may be pictured as a pair of closed loops which consist of joined
straight segments.  The segments join together at {\it kinks}\/ where 
${\bf a}'(u)$ and ${\bf b}'(v)$ are discontinuous. 

 Following
the notation of Ref.~\cite{AC} we take the $a$- and $b$-loops to 
have $N_a$ and $N_b$ linear segments, respectively.
The coordinate $u$ on the $a$-loop is chosen to take the value zero at
one of the kinks and increases along the loop.  The kinks are labeled
by the index $i$ where $i=0,1,\dots,N_a-1$.  The value of $u$ at the
$i$th
kink is denoted by $u_i$ and without loss of generality we set
$u_0=0$.
The segments on the loop are also labeled by $i$, with the $i$th
segment being the one lying between the $i$th and $(i+1)$th kink. The
kink at $u=u_{N_a}$ is the same as the first kink at $u=u_0=0$ but,
even though $u_0$ and $u_{N_a}$ are at the same position on the loop,
$u_0=0$ while $u_{N_a}=1$.  The loop is extended to all values of $u$
by periodicity (with period 1).  We denote ${\bf a}_i = {\bf a}(u_i)$,
and the constant unit vector tangent to the $i$th segment by ${\bf
a}'_i$. Then we have
\begin{equation}
{\bf a}(u) = {\bf a}_i + {\bf a}_i' (u - u_i)  \qquad \hbox{for \ }
u\in[u_i,u_{i+1}] ,
\end{equation}
and for consistency
\begin{equation}
{\bf a}_{i+1} = {\bf a}_i + {\bf a}_i' (u_{i+1} - u_i)  .
\end{equation}
We have corresponding definitions for the ${\bf b}$-loop and we follow the
convention of  Ref.~\cite{AC} by labeling the kinks by the index $j$.

It is now elementary to calculate that, for $n \neq 0$,
\begin{eqnarray}
   {\bf I}^{(n)} &=& {1 \over 2 \pi i n }
       \sum\limits_{i=0}^{N_a-1} 
{ {\bf a}_i' \over 1 - \hat {\bf  \Omega}{\cdot}{\bf a}_i'}
    \left\{ e^{-2\pi i n (u_{i+1} - \hat {\bf  \Omega}{\cdot}{\bf a}_{i+1})} -
 e^{-2\pi i n (u_{i} - \hat {\bf \Omega}{\cdot}{\bf a}_{i})} \right\}
 \nonumber \\
   &=& {i \over 2 \pi  n } \sum\limits_{i=0}^{N_a-1}
      \left\{ { {\bf a}_i' \over 1 - \hat {\bf \Omega}{\cdot}{\bf
          a}_i'} - { {\bf a}_{i-1}' \over 1 - \hat {\bf
          \Omega}{\cdot}{\bf a}_{i-1}'} \right\}
                   e^{-2\pi i n (u_{i} - \hat {\bf \Omega}{\cdot}{\bf a}_{i})} ,
\label{piecewise_linear_I}
\end{eqnarray}
with a similar equation for ${\bf J}$.
If we insert these expressions into Eq. (\ref{As}) and then
into Eq. (\ref{waveformsum}) the sum over $n$ for $h_{R/L}^{\rm rad}$
consists of terms of the form
\begin{equation} 
  \sum_{n=1}^\infty
   { 1 \over n^2} \cos{2 \pi   n [ 2 (t-r) + (u_{i} - \hat {\bf
   \Omega}{\cdot}{\bf a}_{i}) + (v_{j} - \hat {\bf \Omega}{\cdot}{\bf b}_{j})]}
\end{equation}
which may be performed exactly using the identity
\begin{equation}
 \sum_{n=1}^\infty { 1 \over n^2} \cos{2 \pi   n
   x} = \pi^2 (x^2 - x + {\textstyle {1 \over 6}}) \qquad x \in   [0,1) .
\label{identity}
\end{equation}
This function is extended to other values by periodicity, for example,
for $x \in [1,2)$ we merely replace $x$ by $x- 1$ in
Eq.~(\ref{identity}). Such transformations leave the coefficient of
$x^2$ unchanged and can only change the coefficient of $x$ by a multiple
of 2.  As a result when the sum in Eq.~(\ref{piecewise_linear_I}) is 
performed for the coefficient of $x^2$ the sum telescopes and
gives zero.  Thus, {\it the waveform of a piecewise linear loop will be a
piecewise linear function}.  In addition, considering the coefficient of
$x$ all slopes of the waveform must be a multiple of some fundamental 
 slope.  
The slope only changes when a (4-dimensional) kink crosses the past
light cone of the observer at $(t,{\bf x})$. 
These properties are illustrated in the examples below.

\subsection{Garfinkle-Vachaspati loops}
As our first set of loops we study the loops considered by 
Garfinkle and Vachaspati \cite{GV}. The vectors  
${\bf a}(u)$ and  ${\bf b}(v)$ lie in a plane and make a constant
angle  $2 \alpha$ with each other where $\alpha \in (0,\pi/2)$.  
To be specific, we may take ${\bf a}(u)$ and ${\bf b}(v)$ to be   given by
\begin{mathletters}
\begin{equation}
  {\bf a}(u) = \cases{ u (\cos \alpha {\bf i } + \sin \alpha {\bf j}), 
        & $u \in (0, {1 \over 2})$ \cr &\cr
                       (1 - u )(\cos \alpha {\bf i } + \sin \alpha  {\bf j}),
  & $u \in ( {1 \over 2}, 1)$ }
\end{equation}
\begin{equation}
  {\bf b}(v) = \cases{ v  (\cos \alpha {\bf i } - \sin \alpha {\bf j}),
            & $v \in (0, {1 \over 2})$ \cr &\cr
                (1 -v)  (\cos \alpha {\bf i } - \sin \alpha {\bf j}),
            & $v \in ( {1 \over 2}, 1)$. }
\end{equation}
\end{mathletters}
It is then straightforward to calculate that, for $n \neq 0$,
\begin{mathletters}
\begin{eqnarray}
{\bf I}^{(n)} &=&  {e^{i\pi  n \bigl(1 - \sin \theta \cos(\phi -
\alpha)\bigr)} - 1  \over
i \pi n \bigl(1 - \sin^2 \theta \cos^2 ( \phi - \alpha) \bigr)} 
(\cos \alpha {\bf i } + \sin \alpha {\bf j})\\
{\bf J}^{(n)} &=&  {e^{i\pi  n \bigl(1 - \sin \theta \cos(\phi +
\alpha)\bigr)} - 1  \over
i \pi n \bigl(1 - \sin^2 \theta \cos^2 ( \phi + \alpha) \bigr)} 
(\cos \alpha {\bf i } - \sin \alpha {\bf j})
\end{eqnarray}
\end{mathletters}
and correspondingly
\begin{mathletters}
\begin{eqnarray}
A_+^{(n)}&=&    
{  \sin^2\theta  \cos (\phi + \alpha) \cos(\phi - \alpha) - \cos2\phi
    \over
\bigl(1 - \sin^2 \theta \cos^2 ( \phi - \alpha) \bigr)
\bigl(1 - \sin^2 \theta \cos^2 ( \phi + \alpha) \bigr)} \times \nonumber \\
  &&{1 \over n^2 \pi^2}
(e^{i\pi  n \bigl(1 - \sin \theta \cos(\phi -
\alpha)\bigr)} - 1)(e^{i\pi  n \bigl(1 - \sin \theta \cos(\phi +
\alpha)\bigr)} - 1) \\ 
A_\times^{(n)} &=&    
 { \cos \theta \sin 2\phi  \over
\bigl(1 - \sin^2 \theta \cos^2 ( \phi - \alpha) \bigr)
\bigl(1 - \sin^2 \theta \cos^2 ( \phi + \alpha) \bigr)} 
  \times \nonumber \\   &&{1 \over n^2 \pi^2}
(e^{i\pi  n \bigl(1 - \sin \theta \cos(\phi -
\alpha)\bigr)} - 1)(e^{i\pi  n \bigl(1 - \sin \theta \cos(\phi +
\alpha)\bigr)} - 1)  .
\end{eqnarray}
\label{A_GV}
\end{mathletters}
As described above, the sum over $n$ in Eq. (\ref{waveformsum}) may be
 performed explicitly
to yield a piecewise linear function. For example, $\phi \in
[0,\pi/2)$, $h_+$  is given explicitly by
\begin{eqnarray}
h_+&&=    
    {2 G M \over r} { \sin^2\theta  \cos (\phi + \alpha) \cos(\phi - \alpha) - \cos 2\phi 
    \over
\bigl(1 - \sin^2 \theta \cos^2 ( \phi - \alpha) \bigr)
\bigl(1 - \sin^2 \theta \cos^2 ( \phi + \alpha) \bigr)}    \times \nonumber \\
&&\cases{ (1-\sin\theta\cos(\phi-\alpha))(1-\sin\theta\cos(\phi+\alpha)),
&\cr 
\qquad \hfill  0 \leq (t-r) < 
           {\textstyle {1 \over 4}} \sin\theta
 \bigl(\cos(\phi-\alpha)+\cos(\phi+\alpha)\bigr)&\cr
&\cr 
  -8 (t-r)  +
(1+\sin\theta\cos(\phi+\alpha))(1+\sin\theta\cos(\phi-\alpha)), &\cr 
\qquad  \hfill  {\textstyle {1 \over 4}} \sin\theta \bigl(\cos(\phi-\alpha)+ 
\cos(\phi+\alpha)\bigr) \leq  (t-r) < 
 {\textstyle {1 \over 4}} (1+\sin\theta\cos(\phi+\alpha))&\cr
&\cr
- (1-\sin\theta\cos(\phi-\alpha))(1+\sin\theta\cos(\phi+\alpha)),&\cr 
\qquad \hfill  {\textstyle {1 \over 4}} (1+\sin\theta\cos(\phi+\alpha)) \leq  (t-r)
    <  {\textstyle {1 \over 4}} (1+\sin\theta\cos(\phi-\alpha))&\cr
&\cr
8  (t-r - {\textstyle {1 \over 2}}) +  
        (1-\sin\theta\cos(\phi-\alpha))(1-\sin\theta\cos(\phi+\alpha)),&\cr
\qquad \hfill {\textstyle {1 \over 4}} (1+\sin\theta\cos(\phi-\alpha)) \leq
     (t-r) 
 <  {\textstyle {1 \over 2}}.&\cr
}\nonumber \\
\label{gv_waveform}
\end{eqnarray}  
and the waveforms are periodic in $t$ with period $1 \over 2$.
The intervals are ordered in the given way for our choice of $\phi \in
[0,\pi/2)$. $h_\times$ is obtained simply by replacing the prefactor
by that appropriate to $A_\times$ as is clear from Eq.~(\ref{A_GV}).
To obtain the waveforms for other angles we may note that 
the transformation $\phi \to \phi + \pi$ is equivalent to changing the
sign of $(t-r)$, while the transformation $\phi \to \pi - \phi$ is
equivalent  to changing the sign of $(t-r)$ and sign in front of the
$\sin 2\phi$ term in  the prefactor in $h_\times$.

Note that the apparent singularity in the waveforms in the plane of the loop
($\theta=\pi/2$) at $\phi=\pm \alpha$ and $\phi= \pi \pm \alpha$ is
spurious. This may be seen by noting that the waveform is bounded by
the two constant sections of the piecewise linear curve which take on a
value which tends to zero in this limit.  In fact, the numerator of
the prefactor also vanishes in this limit which ensures that the
amplitude tends to zero at these points and hence that even the time derivatives
(which determine the power) are finite. 
Along the axis $\theta=0$, Eq.~(\ref{gv_waveform}) reduces to 
\begin{equation}
h_{+/\times}(\theta=0)=   - {2 G M   \over r } \cos 2\phi/\sin{ 2\phi}
\cases{-8(t-r) + 1, &$0 \leq t-r < {\textstyle {1 \over 4}}$ \cr
 &\cr 
       8(t-r -{\textstyle {1 \over 2}} ) + 1,    
   & ${\textstyle {1 \over 4}} \leq t-r < {\textstyle {1\over 2}}$.}
\end{equation}

\begin{figure}
\centerline{\epsfig{figure=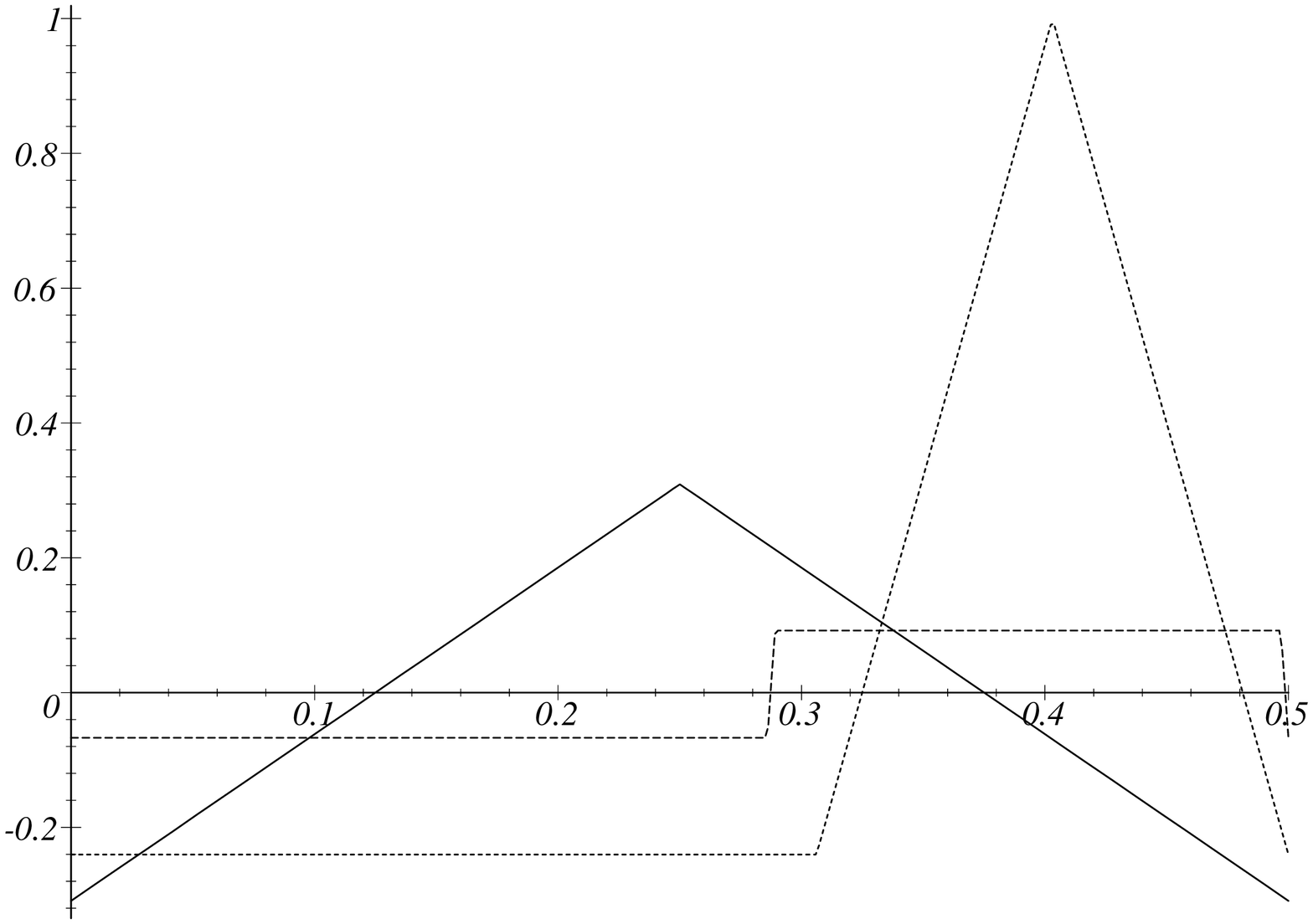,height=12cm,angle=270} }
\caption{Plus-polarized waveforms for the Garfinkle-Vachaspati with
      $\alpha=\pi/4$.
The solid line corresponds to the wave travelling up the $z$-axis ($\theta=0$).
The dotted line corresponds to a direction at elevation $\theta=\pi/3$
along $\phi=0$. The dashed line corresponds to a wave travelling in
the plane of the loop $\theta=\pi/2$) at angle $\phi=\pi/5$.
The cross-polarized waveforms differ only in that there amplitude has
a different dependece on $\theta$ and $\phi$. }
\label{gv_plus}
\end{figure}

Waveforms for various angles a plotted in Fig.~\ref{gv_plus}
for the case of $\alpha=\pi/4$, corresponding to two lines at right
angles.  This is the configuration  which radiates minimum
gravitational  radiation for this class of loops, $P=64 \ln 2\> G\mu^2
\approx 44{\cdot}3614 \> G\mu^2$.

\subsection{Plane-line loops}

As our next set of examples we study the set of loops in which
${\bf a}(u)$ lies along the $z$-axis and ${\bf b}(v)$ is always in the $x$-$y$ plane.
This class of loops was studied by us in Ref.~\cite{ACO} where we gave
an analytic result for the power lost in   gravitational radiation by
such loops.  Explicitly ${\bf a}(u)$ is given by
\begin{equation}
  {\bf a}(u) = \cases{ u {\bf k} & $u \in (0, {1 \over 2})$ \cr &\cr
                       (1 -u) {\bf k} & $u \in ( {1 \over 2}, 1)$. }
\end{equation}
It follows that
 \begin{equation}
{\bf I}^{(n)} =  {e^{i\pi  n (1 -  \cos\theta)} - 1  \over
i \pi n  \sin^2 \theta }  {\bf k} . 
\end{equation}
Also $J^{(n)}_3 =0$, so we have
\begin{mathletters}
\begin{eqnarray}
  A_+ &=&  - \sin\theta \cos \theta 
      ( \cos \phi   J_1   +  \sin \phi J_2  ) I_3 \\
  A_\times &=& \sin \theta ( \sin \phi J_1 - \cos \phi J_2) I_3 .
\label{plane_e}
\end{eqnarray}
\end{mathletters}
It follows immediately that the waveforms vanish along the $z$-axis.

\begin{figure}
\centerline{\epsfig{figure=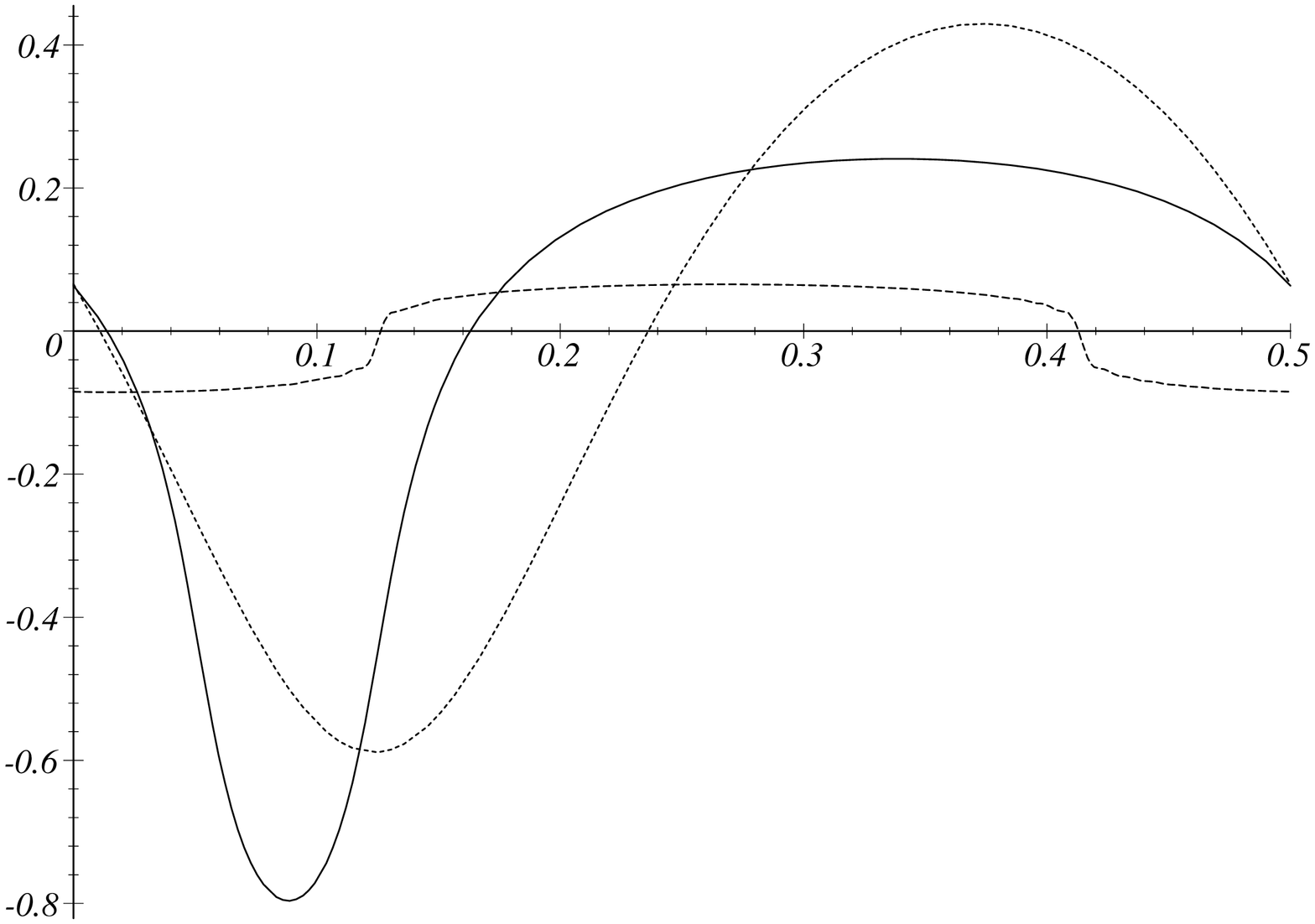,height=12cm,angle=270} }
\caption{Plus-polarized waveforms for the circle-line loop. 
Plotted is $h_+/\sin\theta$ for $\theta = \pi/2$ (solid),
$\theta=\pi/4$
(dashed) and $\theta = \pi/20$ (dotted).   The choice of scaling is
chosen on the basis of the asymptotic form Eq. (\ref{h_plus_asymp}).}
\label{circle_plus}
\end{figure}

\begin{figure}
\centerline{\epsfig{figure=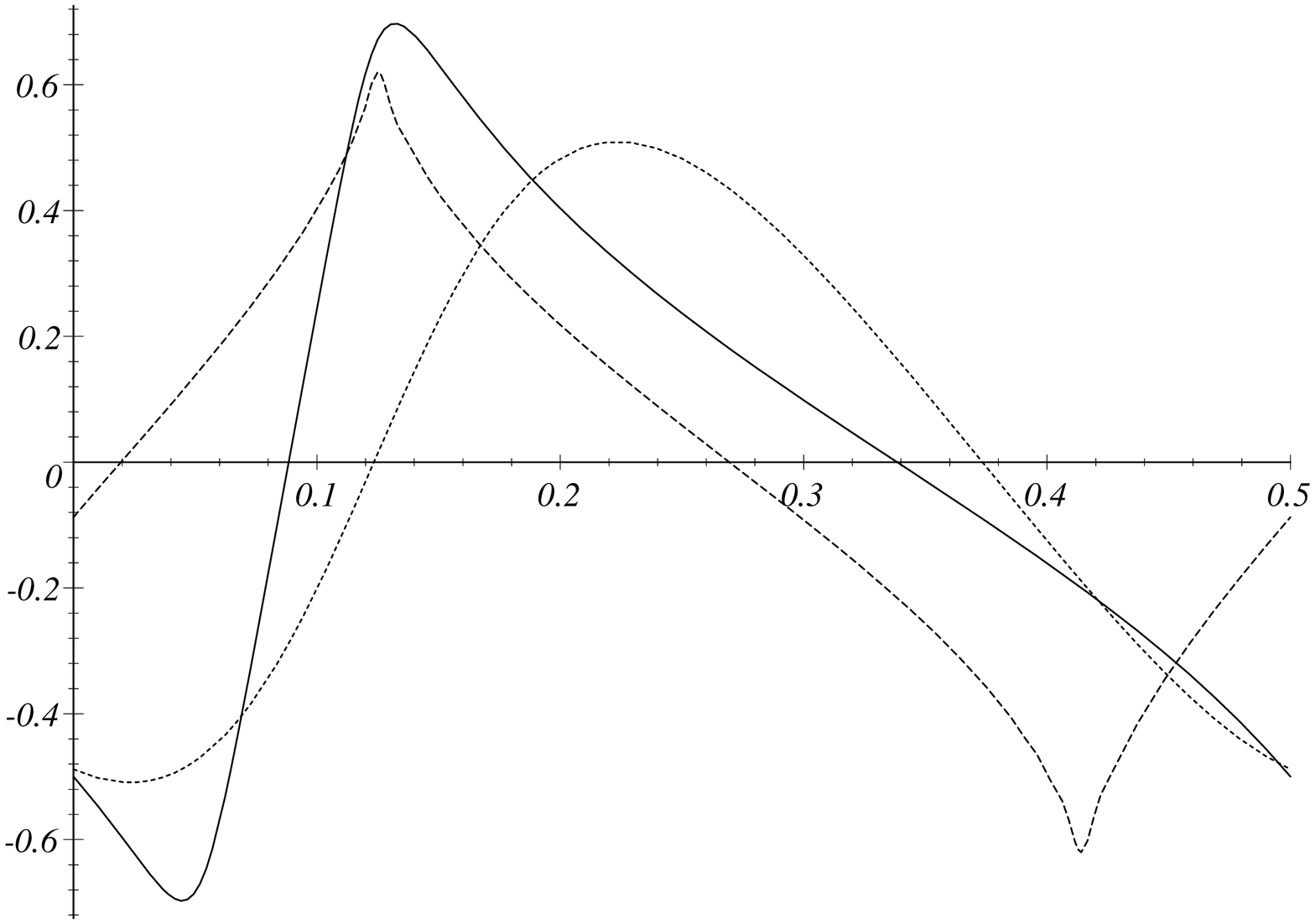,height=12cm,angle=270} }
\caption{Cross-polarized waveforms for the circle-line loop. 
Plotted is $h_\times/\sin\theta$ for $\theta = \pi/2$ (solid),
$\theta=\pi/4$
(dashed) and $\theta = \pi/20$ (dotted).  The choice of scaling is
chosen on the basis of the asymptotic form Eq. (\ref{h_cross_asymp}).}
\label{circle_cross}
\end{figure}

In Ref.~\cite{ACO} we proved that the minimum gravitational radiation
emitted by any loop in this class is given by taking the ${\bf
b}$-loop to be a circle:
 \begin{equation}
{\bf b}(v) = {1 \over 2 \pi} \bigl( \cos (2 \pi  v) {\bf i} +
                     \sin (2 \pi  v) {\bf j} \bigr) .
\end{equation}
The power emitted in gravitational radiation by this loop is
\begin{equation}
  P = 16 \int\limits_0^{2 \pi} {(1 - \cos x) \over x} \, dx \>G \mu^2
  \approx 39{\cdot}0025  \>G \mu^2.
\end{equation} 
${\bf J}^{(n)}$ may be determined explicitly as
\begin{mathletters}
\begin{eqnarray}
J^{(n)}_1 &=& {i \over 2} \left [ e^{i (n+1) (\phi - {\pi \over 2})}
J_{n+1}(n \sin \theta) - e^{i (n-1) (\phi - {\pi \over 2})}
J_{n-1}(n \sin \theta) \right] \\
J^{(n)}_2 &=& {1 \over 2} \left [ e^{i (n+1) (\phi - {\pi \over 2})}
J_{n+1}(n \sin \theta) + e^{i (n-1) (\phi - {\pi \over 2})}
J_{n-1}(n \sin \theta) \right] 
\label{circle_j}
\end{eqnarray}
\end{mathletters}
This gives the equivalent forms
\begin{eqnarray}
  A_+^{(n)} &=& 
  -  {\sin\bigl(\pi n \sin^2(\theta/2)\bigr)  \cos  \theta \over \sin\theta}    
  \left[
    J_{n+1}(n \sin\theta) + 
          J_{n-1}(n \sin\theta) \right] {1 \over \pi
           n} e^{ i \pi n \sin^2(\theta/2) + i n (\phi- \pi/2)}
    \nonumber \\
&=& -  2 
   {\sin\bigl(\pi n \sin^2(\theta/2)\bigr) \cos(\theta) \over \sin^2
     \theta}    J_{n}(n \sin\theta)  {1 \over \pi
           n} e^{ i n \phi - i  n (\pi /2)\cos \theta} .
\end{eqnarray}
and 
\begin{eqnarray}
  A_\times^{(n)} &=& i
    {\sin\bigl(\pi n \sin^2(\theta/2)\bigr)  \over \sin\theta}
  \left[
    J_{n+1}(n \sin\theta) -  J_{n-1}(n \sin\theta) \right] {1 \over \pi
           n} e^{ i \pi n \sin^2(\theta/2) + i n (\phi -
    \pi/2)}  \nonumber  \\
&=& 2i  
   {\sin\bigl(\pi n \sin^2(\theta/2)\bigr) \over \sin \theta} 
            J_n'(n \sin\theta) {1 \over \pi
           n} e^{ i n \phi - i  n (\pi /2)\cos \theta}  .
\end{eqnarray}
The corresponding waveforms for various choices of $\theta$ are
plotted in Figs.~\ref{circle_plus} and \ref{circle_cross}.  (As the
system simply  
rotates cylindrically
with time the choice of $\phi$ is irrelevant, corresponding
simply to a shift in $t-r$.)

In the plane of the ${\bf b}$-loop $h_+$ vanishes so that the wave becomes
linearly polarized.  On the other hand, 
as we approach the axis $\theta=0$ the fundamental mode ($n=1$ term)
dominates and we have
\begin{equation}
\label{h_plus_asymp}
  h_+ \sim  - { GM \over  r  } \sin\theta \sin \bigl(4\pi (t-r)
  + \phi  \bigr)
\end{equation}
and 
\begin{equation}
\label{h_cross_asymp}
  h_\times \sim { GM \over   r  } \sin \theta \cos \bigl(4\pi (t-r)
  + \phi \bigr).
\end{equation}
Thus the wave approaches circular polarization but its amplitude
vanishes as $\sin \theta$. 

\begin{figure}
\centerline{\epsfig{figure=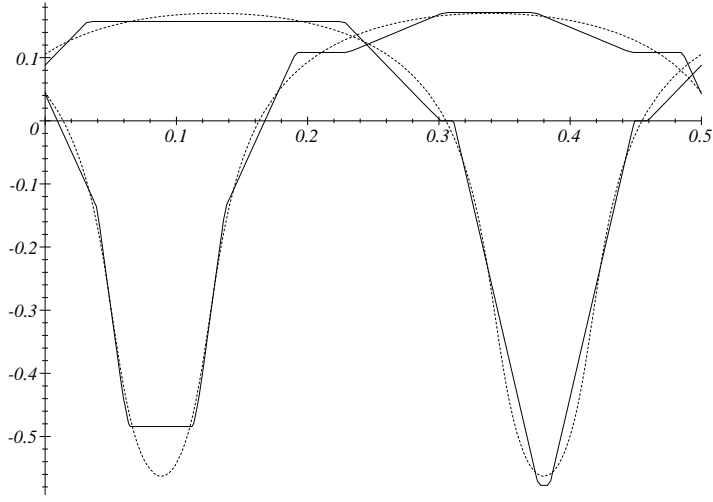,height=12cm,angle=270} }
\caption{The solid lines are the plus-polarized waveforms for the
 hexagon-line loop with $\theta = \pi/4$ and with  
 $\phi =0$ (left through) and  $\phi=5\pi/6$ (right through).
The dotted lines are the corresponding waveforms for the circle-line
loop.}
\label{polygon_plus}
\end{figure}

\begin{figure}
\centerline{\epsfig{figure=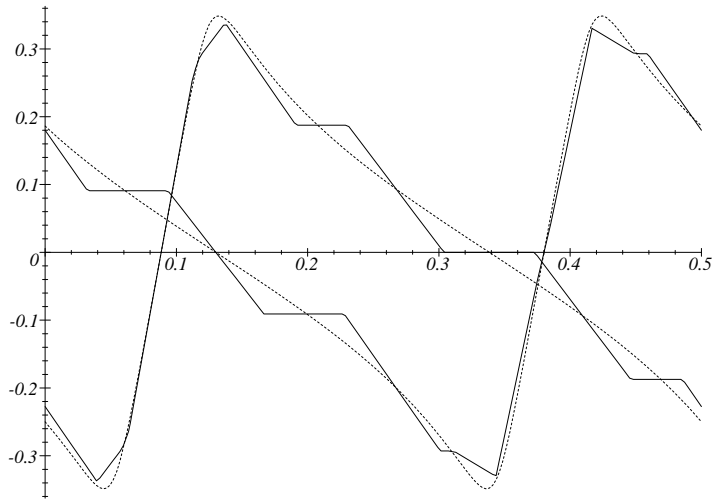,height=12cm,angle=270} }
\caption{The solid lines are the cross-polarized waveforms for the
 hexagon-line loop with $\theta = \pi/4$ and with  
 $\phi =0$ (left peak) and  $\phi=5\pi/6$ (right peak).
The dotted lines are the corresponding waveforms for the circle-line
loop.}
\label{polygon_cross}
\end{figure}

As in   Ref.~\cite{ACO}  we may also consider the case where the ${\bf
b}$-loop forms a
regular $N$-sided polygon.  In Figs.~\ref{polygon_plus} and
\ref{polygon_cross}
we compare the waveform for the circle with that for a regular hexagon
for which 
$P=(40 \ln 5 - 32 \ln 2)\> G\mu^2 \approx 42{\cdot}1968 \> G\mu^2$.
As mentioned above a change in $\phi$ for the circle-line loop
corresponds simply to a shift in $t$, however, this is no longer the
case for the polygon for which the waveform will only repeat every
$2\pi/N$.  Hence in Figs.~\ref{polygon_plus} and
\ref{polygon_cross} we include hexagon-line waveform for both $\phi=0$
and $\phi = 5\pi/6$  (this choice was made simply to disentangle the
two graphs as far as possible).  It is remarkable that even for such a
crude approximation to the circle as a hexagon, the waveform 
of the hexagon-line loop provides remarkably good piecewise
linear approximations to the circle-line waveforms.

\section{CONCLUSION}

\begin{figure}
\centerline{\epsfig{figure=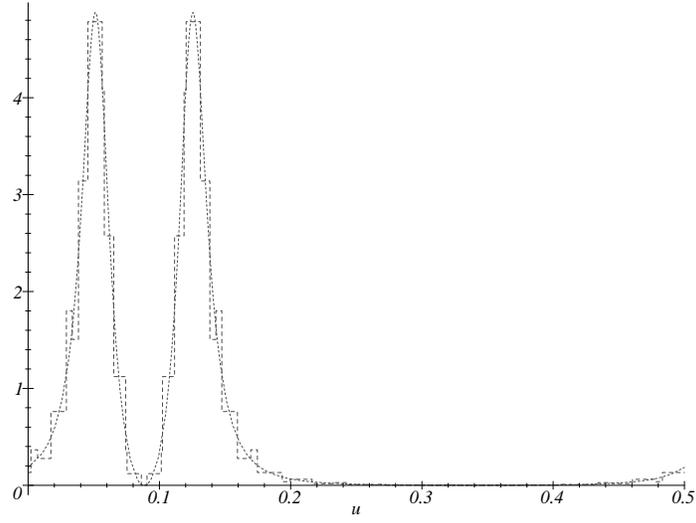,height=12cm,angle=270} }
\caption{Comparison of the `instantaneous power'  in  
plus-polarized waves for the 24-sided polygon-line loop (dotted line)
and for the circle-line  loop (solid line) with $\theta = \pi/4$ 
and $\phi=0$.}
\label{power_fig}
\end{figure}

Given the remarkable agreement of the waveforms it is of interest to
compare the `instantaneous power' defined by 
\begin{equation}
P_{+/\times} = {GM^2 \over 2 \pi} \sum\limits_{n=1}^\infty  
       \omega_n^2 |A_{+/\times}^{(n)}|^2
\end{equation}
in the different polarizations.  While this quantity is not gauge
invariant its time average is and gives the total power radiated in
each polarization. By comparing the function for
the polygon-line loops with the circle-line loop we can certainly see 
that their time averages agree well. 
As the waveform for a piecewise linear loop is a piecewise linear
function, the instantaneous power, which is the square of its
derivative, will be piecewise constant. For example, in
Fig.~\ref{power_fig}
we compare the `instantaneous power' in the plus-polarization between 
the circle-line loop and a regular 24-sided polygon-line loop.
The very close agreement between the two curves provides further
evidence for the validity of the piecewise
linear approximation of string loops used by \cite{AC}.

%%%%%%%%%%%%%%%%%%%%%%%%%%%%%%%%%%%%%%%%%%%%%%%%%%%%%%%%%%%%%%%%%%%%%%%%%%%%%%%

%%%%%%%%%%%%%%%%%%%%%%%%%%%%%%%%%%%%%%%%%%%%%%%%%%%%%%%%%%%%%%%%%%%%%%%%%%%%%%%

\newpage 

\newpage 

%%%%%%%%%%%%%%%%%%%%%%%%%%%%%%%%%%%%%%%%%%%%%%%%%%%%%%%%%%%%%%%%%%%%%%%%%%%%%%%

\end{document}